\documentclass{article} 
\usepackage{nips14submit_e,times}
\usepackage{hyperref}
\usepackage{url}
\usepackage{algorithmicx}
\usepackage{algpseudocode}

\usepackage[pdftex]{graphicx}
\graphicspath{{./},{./figs/}}
\DeclareGraphicsExtensions{.pdf,.jpeg,.png}
\usepackage{caption}
\usepackage{subcaption}

\usepackage{amsfonts}
\usepackage{amsmath}
\usepackage{amsthm}
\usepackage[psamsfonts]{amssymb}
\usepackage{nccmath}
\usepackage{mathtools} 
\usepackage{relsize} 
\usepackage{nicefrac}
\usepackage{times}
\usepackage{bm}
\usepackage{bbm}
\usepackage{verbatim}
\usepackage{cancel}
\usepackage{verbatim}
\usepackage{color}
\usepackage{todonotes}
\usepackage{url}
\usepackage{epsfig}
\usepackage{mathtools}

\usepackage{tikz}
\usetikzlibrary{arrows}

\global\long\def\outputIndex{j}
\global\long\def\dataIndex{i}

\global\long\def\dataDim{p}
\global\long\def\latentDim{q}

\global\long\def\numData{n}

\global\long\def\numInducing{m}

\global\long\def\dataScalar{y}
\global\long\def\dataVector{\mathbf{\dataScalar}}
\global\long\def\dataMatrix{\mathbf{\MakeUppercase{\dataScalar}}}

\global\long\def\latentScalar{x}
\global\long\def\latentMatrix{\mathbf{\MakeUppercase{\latentScalar}}}
\global\long\def\latentVector{\mathbf{\latentScalar}}

\global\long\def\kernelScalar{k}

\global\long\def\kernelMatrix{\mathbf{\MakeUppercase{\kernelScalar}}}

\global\long\def\numData{n}

\global\long\def\dataDim{p}

\global\long\def\mappingFunction{f}
\global\long\def\mappingFunctionVector{\mathbf{\mappingFunction}}
\global\long\def\mappingFunctionMatrix{\mathbf{\MakeUppercase{\mappingFunction}}}

\global\long\def\pseudotargetScalar{u}
\global\long\def\pseudotargetVector{\mathbf{\pseudotargetScalar}}

\global\long\def\inducingInputScalar{z}
\global\long\def\inducingInputVector{\mathbf{\inducingInputScalar}}
\global\long\def\inducingInputMatrix{\mathbf{\MakeUppercase{\inducingInputScalar}}}

\global\long\def\gaussianSamp#1#2{\mathcal{N}\left(#1,#2\right)}

\global\long\def\det#1{\left|#1\right|}

\global\long\def\tr#1{\text{tr}\left(#1\right)}

\usepackage{color}
\usepackage{verbatim}

\definecolor{brown}{rgb}{0.9,0.59,0.078}
\definecolor{ironsulf}{rgb}{0,0.7,.5}
\definecolor{lightpurple}{rgb}{0.156,0,0.245}

\ifdefined\blackBackground
\definecolor{colorOne}{rgb}{0, 1, 1}
\definecolor{colorTwo}{rgb}{1, 0, 1}
\definecolor{colorThree}{rgb}{1, 1, 0}
\definecolor{colorTwoThree}{rgb}{1, 0, 0}
\definecolor{colorOneThree}{rgb}{0, 1, 0}
\definecolor{colorOneTwo}{rgb}{0, 0, 1}
\else
\definecolor{colorOne}{rgb}{1, 0, 0}
\definecolor{colorTwo}{rgb}{0, 1, 0}
\definecolor{colorThree}{rgb}{0, 0, 1}
\definecolor{colorTwoThree}{rgb}{0, 1, 1}
\definecolor{colorOneThree}{rgb}{1, 0, 1}
\definecolor{colorOneTwo}{rgb}{1, 1, 0}
\fi

\ifdefined\blackBackground

\else

\fi

\global\long\def\det#1{\left|#1\right|}

\global\long\def\bfmu{\boldsymbol{\mu}}

\global\long\def\bfPsi{\boldsymbol{\Psi}}

\global\long\def\bfPhi{\boldsymbol{\Phi}}

\global\long\def\bftheta{\boldsymbol{\theta}}

\global\long\def\bfS{\mathbf{S}}

\global\long\def\la{\leftarrow}

\global\long\def\T{{\top}}

\global\long\def\cut#1{}

\global\long\def\detail#1{}

\newcommand{\F}{\mathcal{F}}

\newif\ifsubone
\subonetrue 

\newif\iflayersub
\layersubfalse 

\newcommand{\inducingIndex}{\dataIndex}

\ifsubone
  \renewcommand{\dataIndex}{n}
  \renewcommand{\outputIndex}{d}
  
  \renewcommand{\inducingIndex}{m}
\fi

\newcommand{\jn}{\dataIndex}
\newcommand{\jd}{\outputIndex}

\newcommand{\jm}{\inducingIndex}

\newcommand{\eg}{e.g.\ }

\newcommand{\ra}{\right\rangle}

\newcommand{\intd}{\text{d}}

\renewcommand{\la}{\left\langle}
\renewcommand{\numData}{N}
\renewcommand{\dataDim}{D}
\renewcommand{\latentDim}{Q}

\renewcommand{\numInducing}{M}

\ifsubone
  \newcommand{\n}{_\dataIndex}                
  \renewcommand{\d}{_\outputIndex}            
  \newcommand{\nd}{_{\dataIndex,\outputIndex}}    
  \newcommand{\m}{_\inducingIndex}

\else
  \newcommand{\n}{_{\dataIndex,:}}                
  \renewcommand{\d}{_{:,\outputIndex}}            
  \newcommand{\nd}{_{\dataIndex,\outputIndex}}    
  \newcommand{\m}{_{\inducingIndex, :}}

\fi

\newcommand{\nN}{\numData}
\newcommand{\nQ}{\latentDim}
\newcommand{\nD}{\dataDim}
\newcommand{\nM}{\numInducing}

\newcommand{\sy}{\dataScalar}
\renewcommand{\sf}{\mappingFunction} 

\newcommand{\su}{\pseudotargetScalar}
\newcommand{\sk}{\kernelScalar}

\newcommand{\vx}{\latentVector}
\newcommand{\vy}{\dataVector}
\newcommand{\vf}{\mappingFunctionVector}
\newcommand{\vz}{\inducingInputVector}
\newcommand{\vu}{\pseudotargetVector}

\newcommand{\mX}{\latentMatrix}
\newcommand{\mY}{\dataMatrix}
\newcommand{\mF}{\mappingFunctionMatrix}
\newcommand{\mZ}{\inducingInputMatrix}

\newcommand{\mK}{\kernelMatrix}
\newcommand{\mKnn}{\mK_{\sf\sf}}
\newcommand{\mKnm}{\mK_{\sf\su}}

\newcommand{\mKmm}{\mK_{\su\su}}

\newcommand{\xn}{\vx\n} 

\newcommand{\yn}{\vy\n} 
\newcommand{\yd}{\vy\d} 
\newcommand{\ynd}{\sy\nd}

\newcommand{\um}{\vu\m} 

\newcommand{\fd}{\vf\d} 

\newcommand{\zm}{\vz\m}

\newcommand{\inv}{^{{\mathsmaller{-1}}}}

\iflayersub
  \renewcommand{\dataIndex}{n}
  \renewcommand{\outputIndex}{d}
  
  \renewcommand{\inducingIndex}{m}
  \renewcommand{\n}{^{(\dataIndex)}}                
  \renewcommand{\d}{^{(\outputIndex)}}              
  \renewcommand{\nd}{^{(\dataIndex,\outputIndex)}}  
  \renewcommand{\m}{^{(\inducingIndex)}} 
   
  \renewcommand{\jn}{\dataIndex}
  \renewcommand{\jd}{\outputIndex}
  
  \renewcommand{\jm}{\inducingIndex}
  
  ----------------------------------------------------------------------------------%
\fi

\title{Gaussian Process Models with Parallelization and GPU acceleration}

\author{
Zhenwen Dai\thanks{also at Sheffield Institute for Translational Neuroscience, SITraN.} \qquad Andreas Damianou$^{*}$ \qquad James Hensman$^{*}$ \qquad Neil Lawrence$^{*}$ \\
Department of Computer Science\\
University of Sheffield\\
\texttt{\{z.dai, andreas.damianou, j.hensman, n.lawrence\}@sheffield.ac.uk} \\
}

\nipsfinalcopy 

\begin{document}

\maketitle

\begin{abstract}
In this work, we present an extension of Gaussian process (GP) models with sophisticated parallelization and GPU acceleration. The parallelization scheme arises naturally from the modular computational structure w.r.t.\ datapoints in the sparse Gaussian process formulation. Additionally, the computational bottleneck is implemented with GPU acceleration for further speed up. Combining both techniques allows applying Gaussian process models to millions of datapoints. The efficiency of our algorithm is demonstrated with a synthetic dataset. Its source code has been integrated into our popular software library GPy.
\end{abstract}

\section{Introduction} 
Gaussian processes (GPs), as non-parametric, data driven approaches, are very popular for regression and dimension reduction problems. Their formulation is responsible for their power but also their memory and computational limitations. Considering a regression problem with observed input and output data, collected by rows in matrices $\mY \in \Re^{\nN \times \nD}$ and $\mX \in \Re^{\nN \times \nQ}$ respectively, the input and output are related according to a set of latent functions $f_d$ plus Gaussian noise:
\begin{equation}
\label{eq:generative}
\sy\nd = \sf\d (\xn) + \epsilon\nd,  \epsilon\nd \sim \gaussianSamp{0}{\beta\inv},
\end{equation}
where $\sy\nd$ denotes the $\jd$th dimension of the $\jn$th output point and $\xn$ denotes the $\jn$th input point. A GP prior with covariance function $\sk_\sf$ is placed on the noise-free observation $\sf\d (\xn)$, which can later be integrated out. This will couple the data points in a $\nN \times \nN$ covariance matrix $\mKnn = \sk_\sf(\mX, \mX)$, the inversion of which scales with $\mathcal{O}(\nN^3)$. The main line of work in the literature attempting to speed up GPs is related to low rank approximations \cite{Csato:sparse02,Seeger:fast03,Snelson:pseudo05,Quinonero:unifying05,Lawrence:ivm02,Titsias:variational09} which decouple the function instantiations $\mF = \sf(\mX)$ given a set of $\nM$ auxiliary (or \emph{inducing}) function input-outputs pairs, denoted by $\zm$ and $\um$ respectively, that result in a low rank approximation of $\mKnn$. 
However, even if the resulting computational cost is $\mathcal{O}(\nN \nM^2)$, in big data domain (millions of data points), the computational bottlenecks encountered in practice are actually associated with large $\nN$, since the number of inducing points $\nM$ can be kept small.

In this paper we are interested in scaling up inference of GP models using parallel computations with respect to the data points. To achieve this, we exploit the independence assumptions induced by using auxiliary variables as well as the full independence assumption in $p(\ynd|\fd(\xn))$ of the noise model of equation \eqref{eq:generative}. In practice, this means that all operations involving data points can be written as sums over $\nN$. This observation has recently been exploited by Hensman et al.\  \cite{Hensman:bigdata13} for stochastic variational inference and by Gal et al.\ \cite{Gal:distributed14} for distributed variational inference. Here we adopt a similar formulation, but focus on presenting a distributed implementation which combines large-scale parallelization (with MPI) and GPU acceleration. It is transparently embedded for sparse GP based models in GPy, such as Bayesian GP-LVM \cite{Titsias:bayesGPLVM10,damianou:2014variationalArXiv}, MRD \cite{Damianou:manifold12} and deep GPs \cite{Damianou:deepgp13}.
With experiments on synthetic data, we show that our inference algorithm can efficiently make use of hundreds of computer nodes, which allows for the consideration of genuine big data.

\section{Distributed inference in GP models}

The sparse inference method employed in GPy follows the variational formulation of \cite{Titsias:variational09}, which constructs a variational lower bound to the true log.\ marginal likelihood of the GP using inducing point representations. Specifically, we have, $\log p(\mY | \mX) \ge \sum_{\jd=1}^\nD \F\d$, with:
\begin{equation}\small
\label{eq:abc}
\begin{aligned}
\mathcal{F}\d 
 &=  
	\log 
		\frac{\beta^{\nicefrac{\nN}{2}} 
			  \det{\mKmm}^{\nicefrac{1}{2}}
			 }
			 {(2\pi)^{\nicefrac{\nN}{2}}
			  \det{\mKmm+\beta \bfPhi}^{\nicefrac{1}{2}}
			 }
   - \frac{\beta}{2} \yd^\T \yd \\
 & + \frac{\beta^2}{2} \yd^\T \mKnm(\beta \bfPhi + \mKmm)^{-1} \mKnm^\T \yd 
   -\frac{\beta \phi}{2} + \frac{\beta}{2} \tr{\mKmm^{-1}\bfPhi},
\end{aligned}
\end{equation}
where $\phi = \tr{\mKnn}$ and $\bfPhi = \mKnm^\T \mKnm $. Here, $\mKnm$ contains the cross-covariances between the training and inducing inputs ($\mX$ and $\mZ$) and $\mKmm$ is the matrix of (co)variances between points in $\mZ$. 

As can be seen, the variational bound is factorised with respect to data dimensions. However, for many practical applications we would like parallelization with respect to the number of data, $\nN$.
Towards this direction, we can use the fact that the covariance matrices can be calculated in a decomposable way, as mentioned in \cite{Titsias:bayesGPLVM10}. Specifically, we can write $\phi = \sum_{\jn=1}^\nN \sk_\sf(\xn,\xn)$ and $\bfPhi = \sum_{\jn=1}^\nN (\mKnm)\n^\T(\mKnm)\n $, where $(\mKnm)\n$ is an $\nM-$dimensional vector with its $\jm$th element given by $\sk_\sf(\xn, \zm)$. By additionally introducing $\bfPsi = \sum_{\jn=1}^\nN (\mKnm)^\T\n \yn$, we can re-write the full variational bound (for all dimensions) as $\log(\mY | \mX) \ge \mathcal{F}$, with:
\begin{equation}\small
\label{eq:boundF}
\begin{aligned}
\mathcal{F}
 &=  \nD \log 
		\frac{\beta^{\nicefrac{\nN}{2}} 
			  \det{\mKmm}^{\nicefrac{1}{2}}
			 }
			 {(2\pi)^{\nicefrac{\nN}{2}}
			  \det{\mKmm+\beta \bfPhi}^{\nicefrac{1}{2}}
			 }
   - \frac{\beta}{2} \sum_{\jn=1}^\nN \yn \yn^\T \\
 & + \frac{\beta^2}{2} \bfPsi^\T (\beta \bfPhi + \mKmm)^{-1} \bfPsi
   -\frac{\beta \nD \phi}{2} + \frac{\beta \nD}{2} \tr{\mKmm^{-1}\bfPhi}.
\end{aligned}
\end{equation}
Notice that the above bound is not \emph{fully} factorised, due to the term $(\beta \bfPhi + \mKmm)^{-1}$. However, that term is an $\nM \times \nM$ matrix which is cheap to invert for the usual choices of $\nM$, whereas all the other expensive computations associated with data points can be parallelised.  
In the unsupervised version of this model, namely the Bayesian GP-LVM, $\mX$ is treated as a \emph{latent variable} and is also integrated out, after introducing a prior distribution $p(\mX)$ and a variational posterior distribution $q(\mX)$. Both of these distributions are Gaussian and factorised with respect to $\nN$. Then, similarly to the sparse GP regression case, we can define a variational lower bound on $\log p(\mY)$ where:
\begin{equation}\small
\label{eq:boundBGPLVM}
\log p(\mY) \ge \la \mathcal{F} \ra_{q(\mX)} - \sum_{\jn=1}^\nN \int q(\xn) \log \frac{p(\xn)}{q(\xn)} \; \intd \xn \,.
\end{equation}
Due to the factorisation of $q(\mX)$ w.r.t.\ datapoints, the summations over $\nN$ are maintained in the unsupervised version. Specifically, the only difference between $\F$ and $\la \F \ra_{q(\mX)}$ is that $\phi, \bfPsi$ and $\bfPhi$ are turned into expectations over $q(\mX)$, \eg $\phi = \sum_{\jn=1}^\nN \int_{\xn} \sk_\sf(\xn, \xn) q(\xn) $ and similarly for $\bfPhi$ and $\bfPsi$.
Compared to the approach taken in \cite{Hensman:bigdata13}, the formulation presented here is not fully factorised but results in a tighter bound and, additionally, extends to the unsupervised case straightforwardly. 
This formulation is similar to the distributed variational inference proposed by Gal et al.\ \cite{Gal:distributed14}.
Using the above variational bound as an objective function means that we have the following parameters to optimise: the kernel parameters $\bftheta$, the noise parameter $\beta$ and the inducing inputs $\mZ$. For the rest of our analysis we assume the more general unsupervised scenario, which also involves the parameters of $\{q(\xn)\}_{\jn=1}^\nN$, namely $\{ \bfmu\n, \bfS\n \}_{\jn=1}^\nN$.



As mentioned above, all the computations w.r.t.\ datapoints are combined according to a couple of summations. Therefore, data parallelism is naturally applicable, so that all datapoints are distributed across computer nodes and every node only performs the computation w.r.t.\ its local portion of data, i.e.\ the computation of $\phi$, $\bfPhi$, $\bfPsi$ (and also the second term of equation \eqref{eq:boundBGPLVM} in the Bayesian GP-LVM case). Although all the computations w.r.t.\ datapoints are parallelizable, to recover the exact lower bound, a few steps of indistributable computations have to be done after collecting the intermediate results from individual nodes, e.g., $(\beta \bfPhi + \mKmm)^{-1}$ in particular. After getting the lower bound, the derivatives w.r.t.\ model parameters can be estimated locally. In principle, with a distributed optimizer, all the local parameters can be determined locally, while the global parameters such as kernel parameters can be determined by synchronizing their gradients among all the computer nodes. However, to make use of the existing optimizers like L-BFGS-B in Scipy, we currently collect the gradients of both local and global parameters into one node, estimate the new parameters according to the chosen optimizer, and spread the new parameters across the rest computer nodes. 

\section{GPU acceleration}

\begin{table}
\small
\renewcommand{\arraystretch}{1.3}
\rule{\linewidth}{.4pt}
\begin{algorithmic}[1]
\caption{The GPU function for computing $\bfPsi$ or $\bfPhi$}\label{tab:alg1}
\For{each inducing input $m$ (each pair $(m_1,m_2)$ for $\bfPhi$)} \Comment{distributed across GPU blocks} 
\For{each datapoint $n$} \Comment{distributed across GPU threads}
\State Compute $\bfPsi_{nm}$ or $\bfPhi^{(n)}_{m_1m_2}$. 
\State Write $\bfPsi_{nm}$ into global GPU memory, when computing $\bfPsi$. 
\State Write $\bfPhi^{(n)}_{m_1m_2}$ into shared local memory, when computing $\bfPhi$. 
\EndFor
\State Sum $\bfPhi^{(n)}_{m_1m_2}$ from all threads and write into global memory, when computing $\bfPhi$.
\EndFor
\end{algorithmic}
\begin{algorithmic}[1]
\caption{Computing the gradients that depend on $\bfPsi$ or $\bfPhi$}\label{tab:alg2}
\Require $\frac{\partial L}{\partial \bfPsi}$ or $\frac{\partial L}{\partial \bfPhi}$
\For{each input dimension $q$}
\State Set $(\frac{\partial L}{\partial Z})_{mq}=0$.
\For{each datapoint $n$} \Comment{distributed across GPU threads}
\For{each inducing input $m$ (each pair $(m_1,m_2)$ for $\bfPhi$)} \Comment{distributed across GPU blocks} 
\State Compute $\frac{\partial \bfPsi_{nm}}{\partial \mu_{nq}}$, $\frac{\partial \bfPsi_{nm}}{\partial S_{nq}}$, $\frac{\partial \bfPsi_{nm}}{\partial Z_{mq}}$, $\frac{\partial \bfPsi_{nm}}{\partial \theta_q}$, or $\frac{\partial \bfPhi^{(n)}_{m_1m_2}}{\partial \mu_{nq}}$, $\frac{\partial \bfPhi^{(n)}_{m_1m_2}}{\partial S_{nq}}$, $\frac{\partial \bfPhi^{(n)}_{m_1m_2}}{\partial Z_{m_1q}}$, $\frac{\partial \bfPhi^{(n)}_{m_1m_2}}{\partial \theta_q}$.
\State Compute the derivatives of $L$ w.r.t.\ parameters by combining $\frac{\partial L}{\partial \bfPsi}$ or $\frac{\partial L}{\partial \bfPhi}$.
\State Add $(\frac{\partial L}{\partial Z})_{nmq}$ into $(\frac{\partial L}{\partial Z})_{mq}$.
\EndFor
\State Sum the intermediate results for $(\frac{\partial L}{\partial \mu})_{nq}$ and $(\frac{\partial L}{\partial S})_{nq}$.
\EndFor
\EndFor
\State Sum the intermediate results for $\frac{\partial L}{\partial \bftheta}$.
\end{algorithmic}
\rule{\linewidth}{.4pt}
\end{table}

The scheme mentioned above dramatically scales up computations by exploiting data parallelism. However, we notice that for some models like the Bayesian GP-LVM, the quantities $\bfPhi$ and $\bfPsi$ constitute the computational bottleneck. Computing these quantities requires to go through all datapoints multiple times, something  which can take more than $99\%$ of inference time for large datasets.
 To further speed up the inference, we make use of GPU acceleration. GPU is a type of specialized computation hardware, which consists of a large number of small processing units (e.g., GTX TITAN black has about 2,880 cores). It is extremely efficient at solving a large number of similar small tasks, which is ideal for our inference algorithm. Therefore, we take advantage of the specialized architecture by shifting the computation of $\bfPhi$ and $\bfPsi$ onto GPU. The key for making full use of its computational power is to properly divide computational workloads for parallelization within GPU. Due to the specific architecture of GPU, there are a couple of constraints for programming, e.g., different computing blocks are not synchronized within a GPU function \footnote{More sophisticated synchronization schemes are supported with new cards, but, to support a wide range of GPU cards, we currently stick to compute capability 2.0.}, and synchronization of writing GPU global memory is very expensive. The optimal division of workloads depends on the specific type of data, e.g. the size difference among $N$, $M$, $D$ and $Q$. Our implementation tries to balance between different choices and gives a generic design, which is suitable for most cases.

In particular, we assign each computing block of GPU a subset of inducing inputs for computing $\bfPsi$ and a subset of inducing input pairs for computing $\bfPhi$. We assign each thread within a computing block a subset of datapoints.  With this division, for computing $\bfPsi$, each thread can write its result directly into the GPU memory, and for $\bfPhi$ the intermediate results w.r.t.\ datapoints are stored in shared local memory; these results are later summed together across different threads and written into the GPU memory (see Table \ref{tab:alg1}). For computing the gradients w.r.t.\ parameters that depend on $\bfPsi$ and $\bfPhi$, the same division of workloads is applied. Due to the different sizes of individual parameters, their gradients are produced at different levels of loops (see Table \ref{tab:alg2}).
On top of GPU acceleration, the parallelism scheme mentioned in the previous section can be easily integrated. The computation of the lower bound can be distributed across multiple GPU cards by assigning each GPU card a subset of datapoints. Then, the computation of $\bfPhi$ and $\bfPsi$ for the local portion is shifted onto individual GPU cards, and the local results are combined as mentioned before. 

\section{Experiments}

\begin{figure}[t]
\vspace{-4mm}
        \centering
        \begin{subfigure}[b]{0.5\textwidth}
	\includegraphics[width=\textwidth]{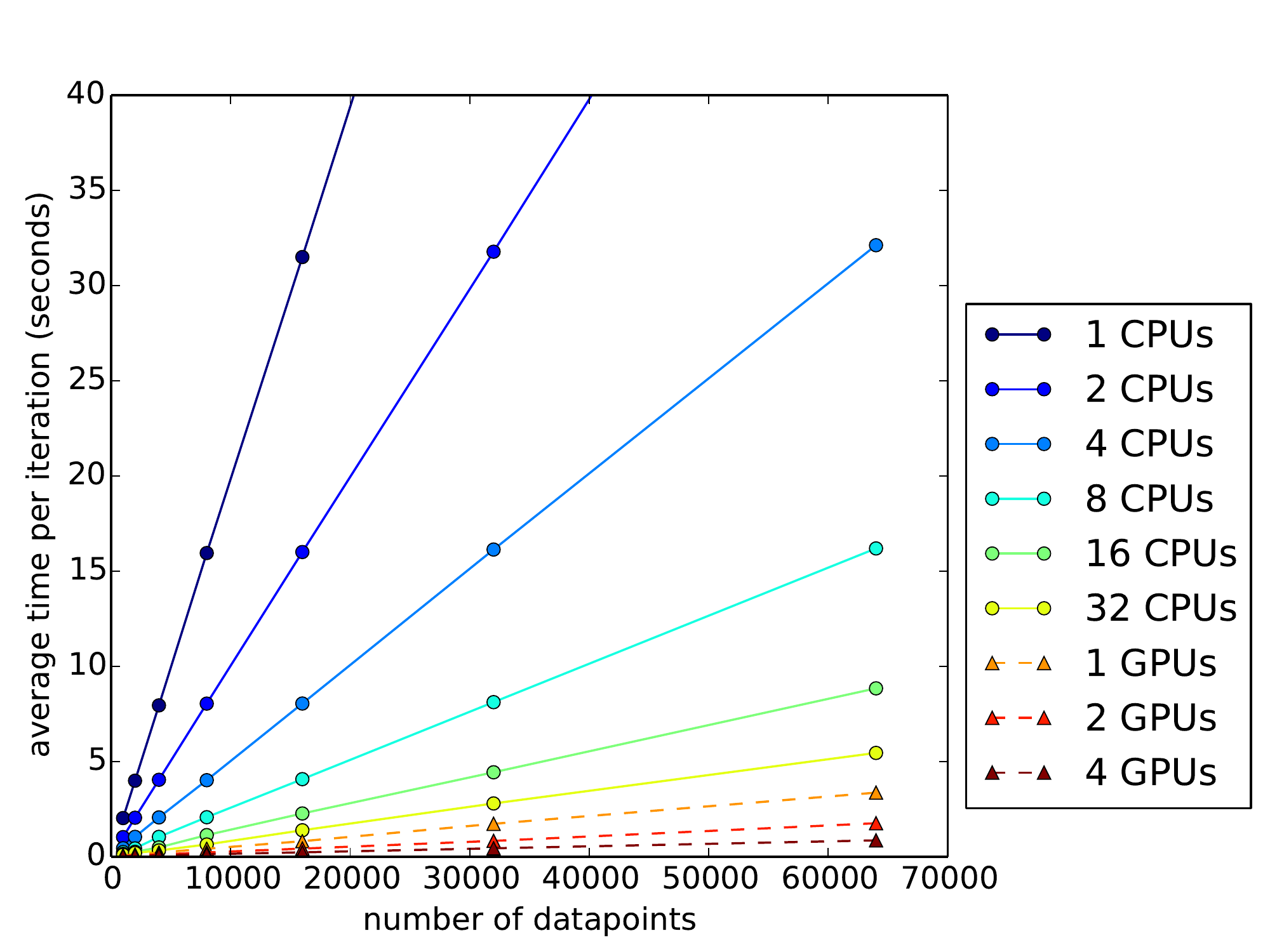}
	\caption{}
	\label{fig:parallel_scaling}
		\vspace{-3mm}
        \end{subfigure}%
        ~ 
        \begin{subfigure}[b]{0.5\textwidth}
	\includegraphics[width=\textwidth]{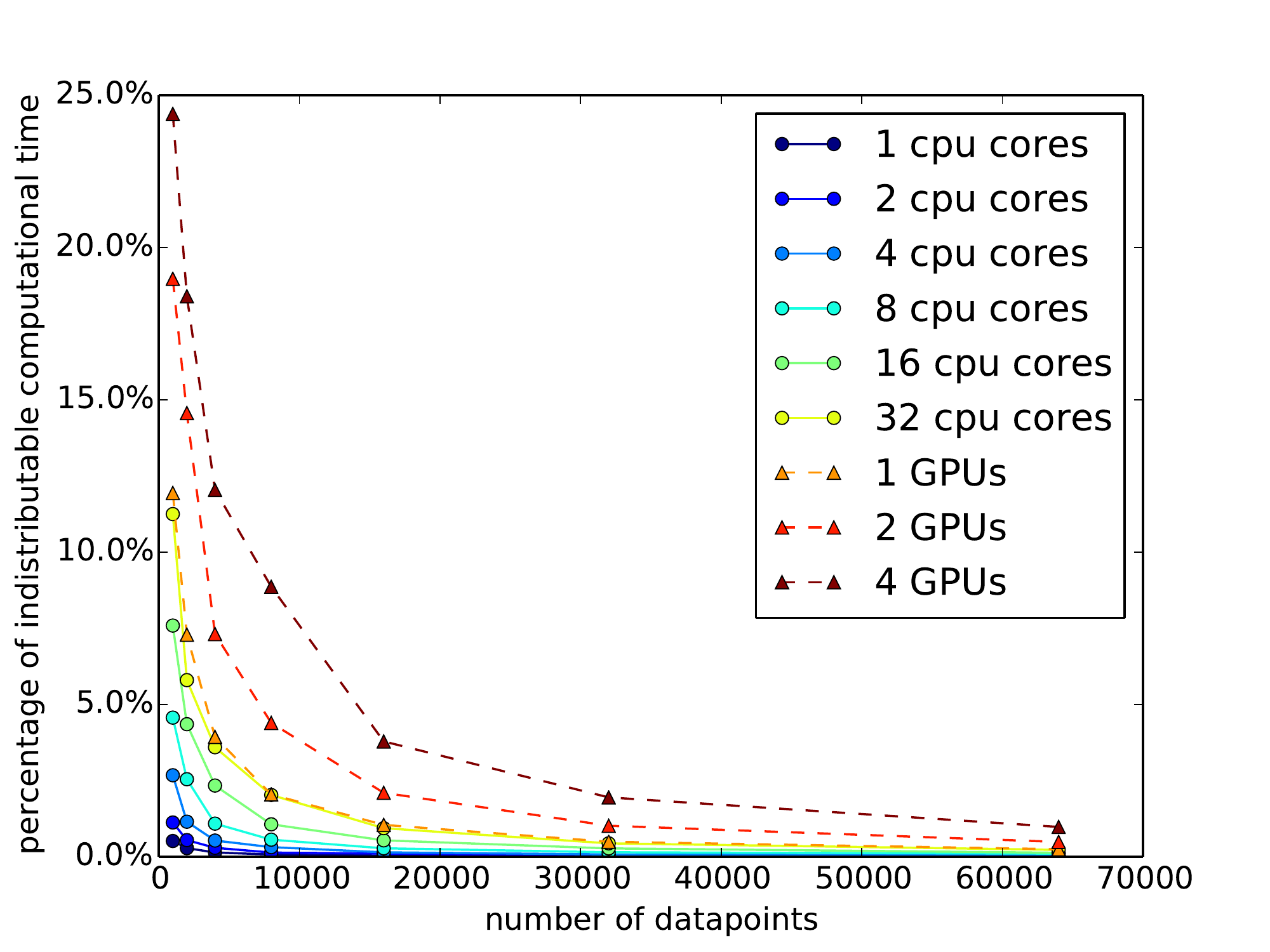}
	\caption{}
	\label{fig:parallel_portion}
	\vspace{-3mm}
        \end{subfigure}
     \caption{(a) The scaling of our parallel inference with both CPUs and GPUs. It shows the average time per iteration for different sizes of datasets. (b) The percentage of time used for indistributable computation per iteration.}
\end{figure}

The performance of our parallelization and GPU acceleration algorithms are evaluated through a synthetic dataset. The dataset was generated by randomly sampling 64k 1D datapoints, and mapping into 3D space by sampling according to a Radius Basis Function (RBF) kernel function. The task is to recover the 1D latent representations from these 3D datapoints by applying the Bayesian GP-LVM as a dimension reduction algorithm. Therefore, the dimensionality of latent space is set to be one, and the number of inducing inputs is set to be 100. A list of datasets with different sizes, ranging from 1k to 64k, are used with a list of different parallelization configurations. All the experiments ran on a single machine with 4 AMD Opteron processors (32 cores in total) and 4 NVIDIA GTX 480 GPU cards, but can directly run on clusters without any changes on the code. 
The efficiency of our parallel inference algorithm is measured by the average computational time per iteration (shown in Fig.\ \ref{fig:parallel_scaling}). The computational time scales linearly with the number of datapoints, which is expected from the computational complexity, and the speed of inference scales roughly linearly with the number of CPUs/GPUs, which shows the efficiency of our parallel implementation, in which the communication overhead is negligible. Note that the speed with a single GPU card is significantly higher than a 32-core computer node. Additionally, the percentage of time used for indistributable computation is shown in Fig.\ \ref{fig:parallel_portion}. It shows that most of time is spent on distributable computation, which means that with more computer resources the speed can further increase. 

\section{Conclusion}
We have presented an efficient parallelised implementation of Gaussian process models and, additionally, the first algorithm with GPU acceleration in this domain. Our results constitute a counter argument against the prejudice that GPs are not applicable to big data, as long as efficient implementations are considered. As future work, we plan to scale up other GP-based models already implemented in GPy and compare our results to methods that currently dominate the domain of big data, like deep learning.

\subsubsection*{Acknowledgments}

We acknowledge funding by the RADIANT and WYSIWYD (EU FP7-ICT) projects. JH was funded by an MRC fellowship.


\bibliographystyle{ieeetr}
\renewcommand*{\refname}{\begin{normalsize}References\end{normalsize}}

\bibliography{../../../bib/other,../../../bib/lawrence,../../../bib/zbooks} 

\end{document}